\documentclass[letterpaper,12pt]{article}

\usepackage{bejournal}          

\usepackage{mathptmx}

\usepackage{graphics}
\usepackage[authoryear,comma,longnamesfirst,sectionbib]{natbib} 
\usepackage[labelfont=bf,labelsep=period,justification=raggedright]{caption}
\usepackage{multirow}
\usepackage{longtable}
\usepackage{booktabs}
\usepackage{epsfig}
\usepackage{setspace}
\usepackage{rotating}
\usepackage{float}
\usepackage{graphicx} \usepackage{epstopdf} \usepackage{float}

\begin{document}


\begin{flushleft}
{\Large
\textbf {Identifying the greatest team and captain $-$ A complex network approach to cricket matches}
}
\\
Satyam Mukherjee$^1$
\\
\bf{1} Department of Chemical and Biological Engineering, Northwestern University, Evanston, Illinois 60208 USA

$\ast$ To whom correspondence should be addressed; E-mail:  satyam.mukherjee@gmail.com.
\end{flushleft}

\section*{Abstract}
We consider all Test matches played between $1877$ and $2010$ and One Day International (ODI) matches played between $1971$ and $2010$. We form directed and weighted networks of teams and also of their captains. The success of a team (or captain) is determined by the `quality' of wins and not on the number of wins alone. We apply the diffusion based PageRank algorithm on the networks to access the importance of wins and rank the teams and captains respectively. Our analysis identifies {\it Australia} as the best team in both forms of cricket $-$ Test and ODI. {\it Steve Waugh} is identified as the best captain in Test cricket and {\it Ricky Ponting} is the best captain in the ODI format. We also compare our ranking scheme with the existing ranking schemes which include the Reliance ICC Ranking. Our method does not depend on  `external' criteria in ranking of teams (captains). The purpose of this paper is to introduce a revised ranking of cricket teams and to quantify the success of the captains.

\section{Introduction}
The study of social networks, representing interactions between humans or groups, is a subject of broad research interest. In recent years, tools from network analysis have been applied to sports. For example, \cite{duch10} developed a network approach to quantify the performance of individual players in soccer. \cite{onody04} studied the complex network structure of Brazilian soccer players. \cite{heuer10} introduced a general model-free approach to elucidate the outcome of a soccer match. Network analysis tools have been applied to football (\cite{girvan02}; \cite{naim05}), baseball (\cite{petersen08}; \cite{sire09}) and basketball (\cite{naim07}; \cite{skinner10}). \cite{saavedra09} studied the head-to-head matchups between Major League Baseball pitchers and batters as a bipartite network (\cite{yellen}). The advantage of a network representation of any real system is that it gives the global view of the entire system and the interaction between individuals reflecting self-emergent phenomena.

In this paper we apply tools of social network analysis to cricket. Cricket is a popular sport around the world and is played mostly in the erstwhile English colonies. Its popularity is the highest in the Indian subcontinent. Despite series of controversies involving match fixing, spot fixing and ball tampering, the sport has managed to maintain international attention as well research interests(\cite{Bailey2004}, \cite{Vani2010}, \cite{Bracewell2009}). Currently there are ten countries that have been granted Test status by International Cricket Council (ICC) - 
Australia (AUS), Bangladesh (BAN), England (ENG), India (IND), New Zealand (NZ), Pakistan (PAK), South Africa (SA), Sri Lanka (SL), West Indies (WI) and Zimbabwe (ZIM). The Reliance ICC Rankings is the official guide used to evaluate the performance of teams as well as the players. Ranking schemes are based on points that are acquired by a team after a tournament. As mentioned by \cite{Vani2010}, due to the opacity of the ranking schemes, the methods used by ICC are still not comprehensible. Again in cricket the captain is responsible for the team. Before the game starts the home captain tosses the coin and the touring captain calls heads or tails. The captain chooses the batting order, sets up fielding positions and shoulders the responsibility of on-field decision-making. Thus the outcome of a match depends on the captain's decisions. Additionally, the captain is also responsible at all times for ensuring that play is conducted within the Spirit of the Game as well as within the Laws \footnote{http://www.lords.org/laws-and-spirit/laws-of-cricket/preamble-to-the-laws,475,ar.html}. In this sense, the success of a team depends on the captain. However, currently there exist no ranking schemes to rank the cricket captains. 

In this paper we numerically estimate  the success of a team as well as the captain by analyzing the network of interaction of competing teams and also the captains. The primary goal of the paper is to elucidate the impact of network structure on rankings of teams and also that of the cricket captains. While the number of wins is a natural measure for success of a team, it does not provide a full picture of the `quality' of win. We are thus motivated to study an alternative method to assess the quality of a win. For example, a win against Australia or South Africa carries more importance than a win against a lesser team.  This is analogous to the citation networks in which the effect of citation coming from an important paper is greater than that coming from a less popular one. The PageRank algorithm (\cite{brin1}), a network-diffusion-based algorithm has emerged as leading method to rank scientists (\cite{radicchi09a}), papers (\cite{chen07}). More recently \cite{radicchi11} applied PageRank algorithm to rank tennis players.  In this paper we apply the PageRank algorithm to rank cricket teams and also identify the most successful cricket captain. The rest of the paper is organized as follows. In Section 2, we define and characterize the cricket-team network and provide a description of the PageRank algorithm that we employ as a ranking scheme across eras and also in the history of cricket ($1877-2010$). In Section 3, we discuss the results and we conclude in Section 4.

\section{Network of Cricket Teams}

\begin{figure}[!ht]
\begin{center}
\includegraphics[width=4in]{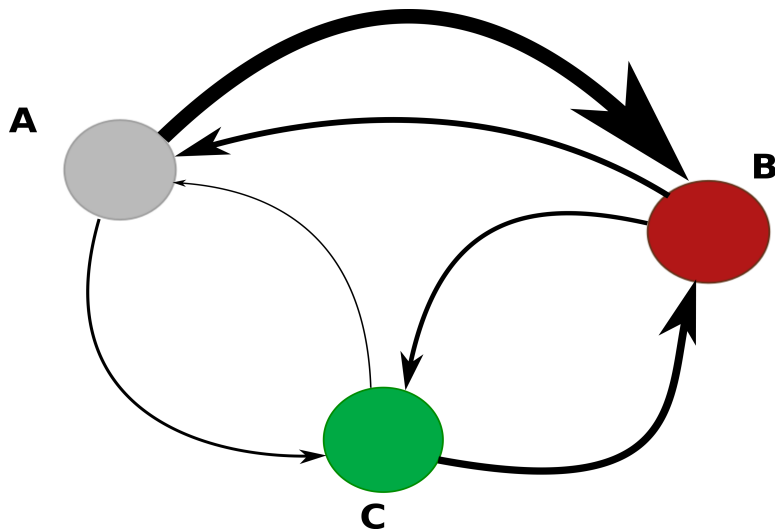}
\end{center}
\caption{{\bf The network of three competing cricket teams.} Three teams {\bf A}, {\bf B} and {\bf C} compete against each other. If {\bf A} defeats {\bf B}, a directed link is established from {\bf B} to {\bf A}. The thickness of the link is proportional to the fraction of wins between {\bf A} and{\bf B}. Thus if we consider all the competing teams a weighted and directed network is established. }
\label{fig1}
\end{figure}
Data were collected from the website of cricinfo ({\it http://www.espncricinfo.com/}). We downloaded the information of results and also the captains who led their respective teams from the score-cards. For a single match, the score-card keeps track of information about the teams, the runs scored by batsmen, wickets taken by bowlers, the names of captains who led their respective teams and the result of a match. We collected the data for Test matches ($1877 - 2010$) and One Day International (ODI) cricket ($1971 - 2010$). In our analysis we have excluded the matches with no results and matches which were abandoned.

We analyze the network of cricket teams by analyzing the head-to-head encounter of competing teams. A single match is represented by a link between two opponents. Thus if team $i$ wins against team $j$, a directed link is drawn from $j$ to $i$ ( Figure~\ref{fig1} ). A weighted representation of the directed network is obtained by assigning a weight $w_{ji}$ to the link, where $w_{ji}$ is equal to the fraction of times team $j$ wins against team $i$. We quantify the relevance of matches with the use of a complex network approach equivalent to the one used for the computation of the PageRank score. Mathematically, the process is described by the following set of equations 

\begin{equation}
    p_i =  \left(1-q\right) \sum_j \, p_j \, \frac{w_{ji}}{s_j^{\textrm{out}}}
+ \frac{q}{N} + \frac{1-q}{N} \sum_j \, \delta \left(s_j^{\textrm{out}}\right) \;\; ,
\label{eq:pg}
\end{equation}
where $w_{ji}$ is the weight of a link and $s_{j}^{out}$ = $\Sigma_{i} {w_{ji}}$ is the out-strength of a link. $p_i$ is the PageRank score assigned to team $i$ and represents the fraction of the overall ``influence'' sitting in the steady state of the diffusion process on vertex $i$ (\cite{radicchi11}).  
In Eqs.~(\ref{eq:pg}), $q \in \left[0,1\right]$ is a control parameter that accounts for the importance of the various terms contributing to the score of the nodes and $N$ is the total number of teams in the network. 
The term $ \left(1-q\right) \, \sum_j \, p_j \, \frac{w_{ji}}{s_j^{\textrm{out}}}$  represents the portion of the score received by node $i$ in the diffusion process according to the hypothesis that vertices  redistribute their entire credit  to neighboring nodes. $\frac{q}{N}$ stands for a uniform redistribution of credit among all nodes. The term $\frac{1-q}{N} \, \sum_j \, p_j \, \delta\left(s_j^{\textrm{out}}\right)$ serves as a correction in the case of the existence of dangling nodes (i.e., nodes with null out-degree), which otherwise would behave as sinks in the diffusion process.


\section{Results}

\begin{figure}[!ht]
\begin{center}
\includegraphics[width=5.5in]{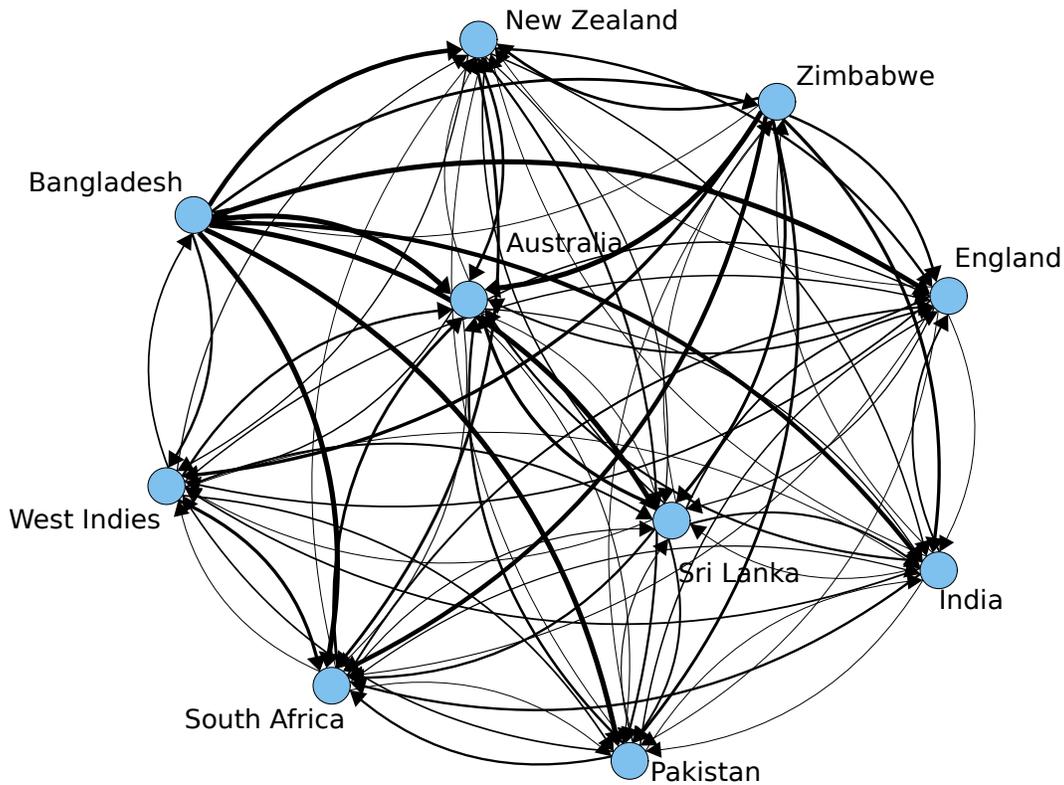}
\end{center}
\caption{ The network of teams in the history of Test cricket ($1877-2010$). }
\label{fig2}
\end{figure}

Traditionally, the choice of $q$ is set at $0.15$ (\cite{brin1}). Hence, we set $q=0.15$ and run the ranking scheme on networks of cricket teams and also on their captains. In Table~\ref{table1}, we report the results obtained from analysis of network of cricket teams for Test cricket. We identify {\it Australia} as the most successful team in history of Test cricket. Even though {\it South Africa} was banned from playing international cricket from $1970 - 1991$, it emerges as the second best team followed by {\it England}, {\it West Indies}, {\it Pakistan}, {\it India}, {\it  Sri Lanka}, {\it New Zealand}, {\it Zimbabwe} and {\it Bangladesh}. Table~\ref{table2} shows the ranking of teams in history of ODI cricket ($1971 - 2010$). Again, {\it Australia} emerges as the best ODI team ever followed by {\it South Africa}, {\it West Indies}, {\it England}, {\it Pakistan}, {\it India}, {\it New Zealand}, {\it  Sri Lanka}, {\it Zimbabwe} and {\it Bangladesh}. The success of Australia could be justified by the dominance of {\it Australia} in International cricket for a long period of time. {\it Australia} won test series in all the countries and also won four ICC World cups in $1987$, $1999$, $2003$ and $2007$.
\newline
\begin{figure}[!ht]
\begin{center}
\includegraphics[width=5.5in]{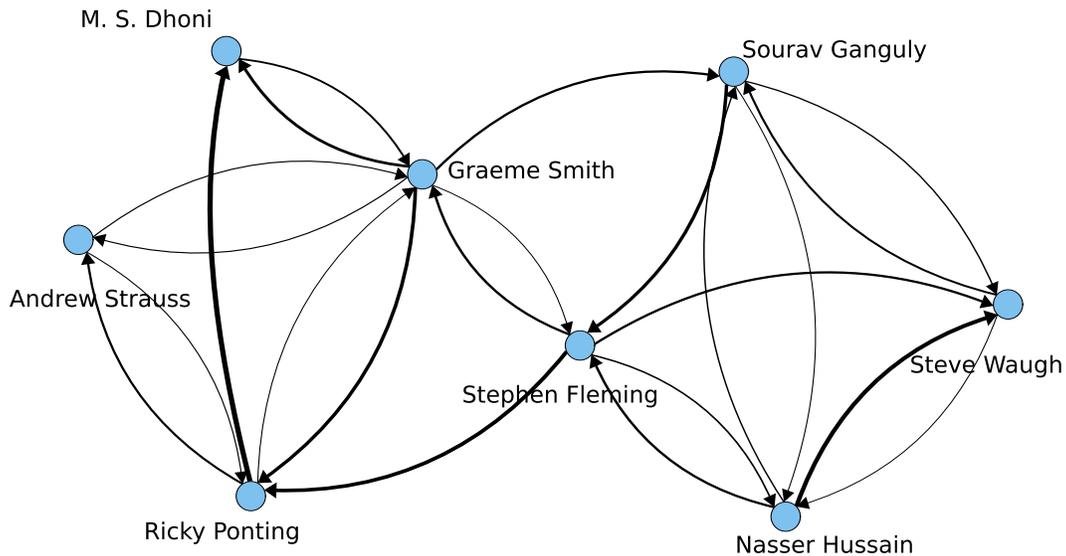}
\end{center}
\caption{ Subraph of the network of most succesful captains in the history of Test cricket ($1877-2010$). }
\label{fig3}
\end{figure}
We also report the results obtained from the analysis of the network of competing captains(See Table~\ref{table3}). {\it Steve Waugh}  heads the top $20$ list of most successful captains in Test cricket. The success of {\it Steve Waugh} could be {\it posteriori} justified by the fact that he led {\it Australia} in $15$ of their world-record $16$ successive Test victories. Over all {\it Steve Waugh} won $72\%$ of the Test matches he captained. It is interesting to note that $8$ of the top $20$ captains are from {\it Australia}. South Africa's {\it Graeme Smith} emerges as the second best captain with {\it Ricky Ponting} occupying the third position. From the subcontinent only India's {\it M. S. Dhoni} and {\it Sourav Ganguly} finds a place in the top $20$ list. We also perform a similar analysis in ODI cricket (See Table~\ref{table4}). This time {\it Ricky Ponting } emerges as the best captain in ODI history, followed by {\it Graeme Smith} (South Africa) in second place and {\it Imran Khan} (Pakistan) in the third. {\it Ricky Ponting}'s success as a captain in the ODI format is marked by two successive World Cup wins in $2003$ and $2007$, with a world-record of $34$ consecutive undefeated World Cup games. Under his captaincy {\it Australia} also won the Champions trophy in $2006$ and successfully defended the title in  $2009$. Contrary to the list in Test cricket, several of the successful captains in the ODI format are from the subcontinent. 
\newline
We also perform a different kind of analysis by constructing networks of teams and their captains in different eras. In Table~\ref{table5a} and Table~\ref{table5b} we report the ranking of teams in different era of Test cricket respectively. We compare our ranking with Reliance ICC Team Ranking\footnote{ The Reliance ICC Team Rankings were launched for ODI cricket in 2002 and for Test cricket in 2003.}. The table of historical ranking of teams, available at ICC's website($http://icc-cricket.yahoo.net/match\_zone/historical\_ranking.php$), begins from $1951$ for Test cricket and from $1981$ for ODI cricket. We rank the teams according to the average of the points scored by any team. 
\newline
During the period $1877-1951$, {\it Australia} emerged as the most successful team. Between $1952$ and $1960$ {\it Australia} was the most successful team according to the PageRank algorithm and also ICC's ranking scheme. During $1961-1970$ {\it West Indies} was the best team according to ICC ranking. Even though the early 1960s were poor periods for {\tt England}, during the late 60's {\it England} defeated stronger opponents like {\it West Indies} and {\it Australia}. Hence judging by the quality of wins, according to PageRank during $1961-1970$ England was the most successful team. A similar effect is also observed  during the $1971-1980$ era, where {\it India} occupies the second position according to PageRank. During the same period {\it India} defeated stronger opponents like {\it West Indies} and {\it England}. 
\newline
Both ranking schemes show {\it West Indies} was the best team between $1981$ and $1990$. Their best period was between February 1981 and December 1989: in $69$ Tests in that span, they had a $40$-$7$ win-loss record, with victories against {\it Australia}, {\it England}, {\it New Zealand} and {\it India}. During the same span, {\tt Pakistan} was victorious against quality opposition like {\it Australia}, {\it England}, and {\it India}. We observe that both ranking schemes predict {\it Australia} as the best team since then. The dominance of {\it Australia} in both decades is also reflected in the fact that between October $1999$ and November $2007$, they played $93$ Tests, and won $72$ of them with $72$-$10$ win-loss record. The ranking of other teams according to PageRank does not correspond to those of ICC Ranking. During $1991-2000$ {\it India} occupies the third position according to PageRank score, instead of {\it West Indies}. Similarly, between $2001$ and $2010$, {\it India} occupies the second position according to PageRank, whereas according to the ICC Ranking {\it South Africa} occupies the second spot. 
\newline
We report a similar ranking of teams in ODI cricket in different era in Table~\ref{table6a}. We observe that {\it West Indies} was the best team throughout the 70's and 80's. PageRank score shows that {\it South Africa} was the best team in the 90's and {\it Australia} is the best team from $2000-2010$. According to ICC Ranking {\it Australia} is the most successful team during the 1990s and also from $2000-2010$. We observe strong correlation between PageRank score and Reliance ICC Ranking and fraction of victories (in-strength rank). We compare the overall ranking of teams playing Test cricket ($1952-2010$) and ODI cricket ($1981-2010$). Figure~\ref{fig4}(a) shows that between $1952$ and $2010$ {\it South Africa} is the best team according to PageRank score, where as {\it Australia} is the best team according to Reliance ICC Ranking. We observe strong correlation between the ranking schemes for ODI cricket ($1981-2010$) (as shown in Figure~\ref{fig4}(b)). According to PageRank score and in-strength the top three positions in Test cricket ($1877-2010$),  are occupied by {\it Australia}, {\it South Africa} and {\it England} respectively (see Figure~\ref{fig4}(c)). In ODI cricket ($1971-2010$), {\it Australia} emerges as the best team according to PageRank score as well as in-strength. In Figure~\ref{fig5} we show the correlation among different ranking schemes as function of time. 
\newline
We provide a ranking of captains in Test cricket (Table~\ref{table5c}) and ODI cricket (Table~\ref{table6b}) in different era. Between $1877$ and $1951$ {\it Bill Woodfull} (Australia) is the most successful captain with {\it Sir Don Bradman} occupying the second position. {\it Richie Benaud} (Australia) leads the list twice during $1952-1960$ and $1961-1970$. During the period $1971-1980$ {\it Ian Chappell} occupies the top position as captain, with {\it Clive Lloyd} occupying the second position. From $1981-1990$ West Indies was the most successful team and {\it Sir Vivian Richards} was the most successful captain. {\tt Mark Taylor} (Australia) is the best captain between $1991$ and $2000$ and {\it Graeme Smith} (South Africa) emerge as the best captain during $2001-2010$. 
In ODI cricket Australia's {\it Greg Chappell} emerge as the most successful captain between $1971$ and $1980$. {\it Clive Lloyd} occupy the second position during that period. Pakistan's {\it Imran Khan} leads the list during the $1981-1990$ era. South Africa's {\it Hansie Cronje} was the most successful captain from $1991-2000$. During the period $2000-2010$ {\it Ricky Ponting} is the most successful captain followed by South Africa's {\it Graeme Smith} and India's {\it M.S.Dhoni}. In Figure~\ref{fig6} we show the correlation among the two ranking schemes for captains.

\begin{table}
\centering
\caption{{\bf Most successful teams in history of Test cricket ($1877 - 2010$}). The teams are ranked according to the PageRank score of each team.}
\begin{tabular}{ll}
\hline
\textbf{Rank}       & \textbf{Team}       \\
\hline
$1$    & Australia      \\
$2$    & South Africa        \\
$3$    & England      \\
$4$    & West Indies      \\
$5$    & Pakistan     \\
$6$    & India  \\
$7$    & Sri Lanka \\
$8$    & New Zealand  \\
$9$    & Zimbabwe  \\
$10$   & Bangladesh  \\
\hline
\end{tabular}
\label{table1}
\end{table}

\begin{table}
\centering
\caption{{\bf Most successful teams in the history of ODI cricket ($1971 - 2010$). } The teams are ranked according to the PageRank score of each team.}
\begin{tabular}{ll}
\hline
\textbf{Rank}        & \textbf{Team}       \\
\hline
$1$    & Australia      \\
$2$    & South Africa        \\
$3$    & West Indies      \\
$4$    & England     \\
$5$    & Pakistan  \\
$6$    & India  \\
$7$    & New Zealand \\
$8$    & Sri Lanka  \\
$9$    & Zimbabwe  \\
$10$   & Bangladesh  \\
\hline
\end{tabular}
\label{table2}
\end{table}


\begin{table}
\centering
\caption{\textbf{Top twenty captains in Test cricket ($1877-2010$). } We also provide the nationality of the captain. The captains are ranked according to the PageRank score of each captain.}
\begin{tabular}{lll}
\hline
\textbf{Rank}        & \textbf{Captain}   & \textbf{Country}    \\
\hline
$1$    & Steve Waugh  & Australia    \\
$2$    & Graeme Smith & South Africa     \\
$3$    & Ricky Ponting & Australia   \\
$4$    & Greg Chappel & Australia   \\
$5$    & Richie Benaud & Australia  \\
$6$    & Clive Lloyd & West Indies \\
$7$    & Ian Chappel & Australia  \\
$8$    & Allan Border & Australia  \\
$9$    & M. S. Dhoni & India  \\
$10$   & Nasser Hussain  & England  \\
$11$    & Peter May  & England  \\
$12$    & Bill Woodfull & Australia     \\
$13$    & Sir Vivian Richards & West Indies  \\
$14$    & Sir Frank Worell & West Indies  \\
$15$    & Sourav Ganguly & India  \\
$16$    & Kim Hughes & Australia   \\
$17$    & Ray Illingworth & England  \\
$18$    & Geoff Howarth & New Zealand \\
$19$    & Andrew Strauss & England  \\
$20$   & Stephen Fleming  & New Zealand  \\

\hline
\end{tabular}
\label{table3}
\end{table}

\begin{table}
\centering
\caption{\textbf{Top twenty captains in ODI cricket ($1971-2010$). }We also provide the nationality of the captain. The captains are ranked according to the PageRank score of each captain.}
\begin{tabular}{lll}
\hline
\textbf{Rank}        & \textbf{Captain}   & \textbf{Country}  \\
\hline
$1$    & Ricky Ponting & Australia      \\
$2$    & Graeme Smith & South Africa        \\
$3$    & Imran Khan & Pakistan \\
$4$    & Hansie Cronje & South Africa     \\
$5$    & Arjuna Ranatunga & Sri Lanka   \\
$6$    & Stephen Fleming & New Zealand  \\
$7$    & Clive Lloyd & West Indies \\
$8$    & M. S. Dhoni & India   \\
$9$    & Sir Vivian Richards & West Indies \\
$10$   & Kapil Dev & India \\
$11$    & Allan Border & Australia \\
$12$    & Mahela Jayawardene & Sri Lanka        \\
$13$    & Brian Lara & West Indies    \\
$14$    & Daniel Vettori & New Zealand    \\
$15$    & Paul Collingwood & England     \\
$16$    & Sourav Ganguly & India \\
$17$    & Mohammad Azharuddin & India \\
$18$    & Rahul Dravid & India \\
$19$    & Javed Miandad & Pakistan \\
$20$   & Wasim Akram & Pakistan \\

\hline
\end{tabular}
\label{table4}
\end{table}

\begin{table}
\centering
\caption{{\bf Ranking of teams in different era in Test history.} We have shown the ranking from $1877 - 1980$. There exist no ICC ranking during $1877-1950$. }
\begin{tabular}{ccc}

\toprule
\multirow{2}{*}{}
 \textbf{Era} & \textbf{PageRank} &\textbf{Reliance ICC-Ranking} \\ 
\midrule 
\multirow{6}{*}{\textbf{1877-1950}} & \multirow{6}{*}{} & \multirow{6}{*}{\textbf{-NA-}}	\\
	& Australia &\\ 
	& England     &   \\
	& West Indies  & \\ 
	& South Africa  &   \\
        & New Zealand   & \\ 
	& India    & \\

\midrule 

\multirow{6}{*}{\textbf{1951-1960}}	
	& \multirow{6}{*}{}
	
	Australia   &  Australia \\ 
	& England     & England   \\
	& Pakistan  & West Indies \\ 
	& West Indies   & South Africa  \\
        & South Africa  & Pakistan \\ 
	& India    & India\\
	& New Zealand  &  New Zealand \\
\midrule 

\multirow{7}{*}{\textbf{1961-1970}}	
	& \multirow{6}{*}{}
	
	England      & West Indies \\ 
	& West Indies   & Australia    \\
	& Australia    &  England  \\
	& New Zealand   & South Africa\\ 
	& South Africa   & India  \\
        & India    & Pakistan \\
	& Pakistan   & New Zealand \\
\midrule 

\multirow{7}{*}{\textbf{1971-1980}}	
	& \multirow{6}{*}{}
	
	Australia   & Australia  \\ 
	& India      & England   \\
	& West Indies  & Pakistan \\ 
	& England    & West Indies \\
        & Pakistan   & India \\ 
	& New Zealand  & New Zealand   \\

\bottomrule
\end{tabular}
\label{table5a}
\end{table}

\begin{table}
\centering
\caption{{\bf Ranking of teams in different era in Test history.} We have shown the ranking from $1981 - 2010$. }
\begin{tabular}{ccc}

\toprule
\multirow{2}{*}{}
 \textbf{Era} & \textbf{PageRank}  &\textbf{Reliance ICC-Ranking}\\ 
\midrule 

\multirow{8}{*}{\textbf{1981-1990}}	
	& \multirow{8}{*}{}
	
	West Indies & West Indies     \\ 
	& Pakistan   & Pakistan     \\
	& Australia  & New Zealand \\ 
	& New Zealand   & Australia  \\
        & England  & India \\ 
	& India    & England\\
	& Sri Lanka   & Sri Lanka \\
	& Zimbabwe   & Zimbabwe \\

\midrule 

\multirow{10}{*}{\textbf{1991-2000}}	
	& \multirow{10}{*}{}
	
	Australia    &  Australia \\ 
	& South Africa   & South Africa    \\
	& India   & West Indies\\ 
	& West Indies  & Pakistan   \\
        & Pakistan   & India \\ 
	& England   & England \\
	& New Zealand  & Sri Lanka \\
	& Sri Lanka  & New Zealand \\
	& Zimbabwe   & Zimbabwe\\
	& Bangladesh  & Bangladesh \\
\midrule 
\multirow{10}{*}{\textbf{2001-2010}}	
	& \multirow{10}{*}{}
	
	Australia   &   Australia \\ 
	& India      &     South Africa \\
	& South Africa  &   India \\
        & England  &      England \\
	& Sri Lanka  &    Sri Lanka\\
	& Pakistan  &     Pakistan\\
        & New Zealand  &  New Zealand \\ 
	& West Indies  &  West Indies \\
	& Zimbabwe     &  Zimbabwe\\
	& Bangladesh    & Bangladesh\\

\bottomrule
\end{tabular}
\label{table5b}
\end{table}

\begin{table}
\centering
\caption{{\bf Ranking of teams in different era in ODI history.} We construct network of teams for each era. The teams are then ranked according to the PageRank score and compared with the Reliance ICC Ranking of Teams. During the period $1981-1990$ Zimbabwe and Bangladesh receive no points in the Reliance ICC Ranking and hence their ranks are not listed.}
\begin{tabular}{ccc}

\toprule
\multirow{2}{*}{}
 \textbf{Era} & \textbf{PageRank} &\textbf{Reliance ICC-Ranking} \\ 
\midrule 
\multirow{8}{*}{\textbf{1971-1980}}	
	& \multirow{8}{*}{}
	& \multirow{8}{*}{\textbf{-NA-}} \\

	& West Indies  & \\ 
	& Australia     &  \\
	& England  & \\ 
	& New Zealand    & \\
        & Pakistan   &\\ 
        & India   &\\ 
        & Sri Lanka  & \\ 

\midrule 

\multirow{8}{*}{\textbf{1981-1990}}	
	& \multirow{8}{*}{}
	
	West Indies     & West Indies \\ 
	& Australia    & Australia     \\
	& England  & England\\ 
	& Pakistan    & Pakistan\\
        & India   & India\\ 
        & New Zealand  & New Zealand\\ 
        & Sri Lanka   & Sri Lanka\\ 
        & Zimbabwe   & $-$\\ 
        & Bangladesh & $-$ \\
\midrule 

\multirow{10}{*}{\textbf{1991-2000}}	
	& \multirow{10}{*}{}
	
	South Africa    & Australia \\ 
	& Australia    & South Africa   \\
	& Pakistan   & Pakistan    \\
	& England  & West Indies\\ 
	& Sri Lanka    & England \\
	& West Indies   & India  \\
	& India   & Sri Lanka \\
	& New Zealand   & New Zealand \\
 	& Zimbabwe    & Zimbabwe\\
        & Bangladesh  & Bangladesh   \\
\midrule 

\multirow{10}{*}{\textbf{2001-2010}}	
	& \multirow{10}{*}{}
	
	Australia     & Australia \\ 
	& South Africa      & South Africa \\
	& India  & Sri Lanka \\ 
	& Sri Lanka    & Pakistan  \\
        & Pakistan   & India \\ 
        & New Zealand    & New Zealand\\ 
        & England   & England \\ 
        & West Indies   & West Indies \\ 
        & Bangladesh   & Zimbabwe \\ 
        & Zimbabwe   & Bangladesh \\ 

\bottomrule
\end{tabular}
\label{table6a}
\end{table}

\begin{figure}[!ht]
\begin{center}
\includegraphics[width=5.5in]{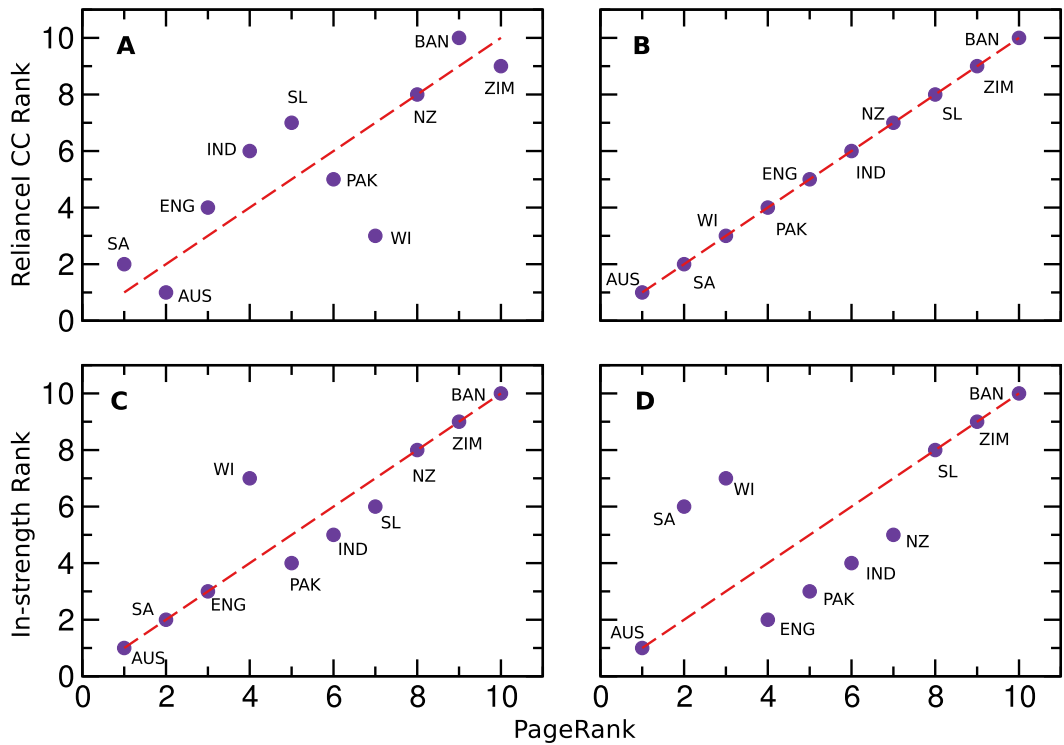}
\end{center}
\caption{{\bf Relation between different ranking schemes. } {\bf (A)} Scatter plot between the rank positions obtained according to Reliance ICC Ranking and those obtained with PageRank for Test cricket ($1952-2010$); (Kendall ${\tau}=0.644$, Spearman correlation ${\rho}=0.818$). {\bf (B)} Scatter plot between the rank positions obtained according to Reliance ICC Ranking and those obtained with PageRank for ODI cricket ($1981-2010$); (${\tau}=1.0$, ${\rho}=1.0$). {\bf (C)} Scatter plot between the rank positions obtained according
to in-strength and those obtained with PageRank for Test cricket ($1877-2010$); (${\tau}=0.867$, ${\rho}=0.927$). {\bf (D)} Scatter plot between the rank positions obtained according
to in-strength and those obtained with PageRank for ODI cricket ($1971-2010$); (${\tau}=0.644$, ${\rho}=0.709$). }
\label{fig4}
\end{figure}

\begin{figure}[!ht]
\begin{center}
\includegraphics[width=5.5in]{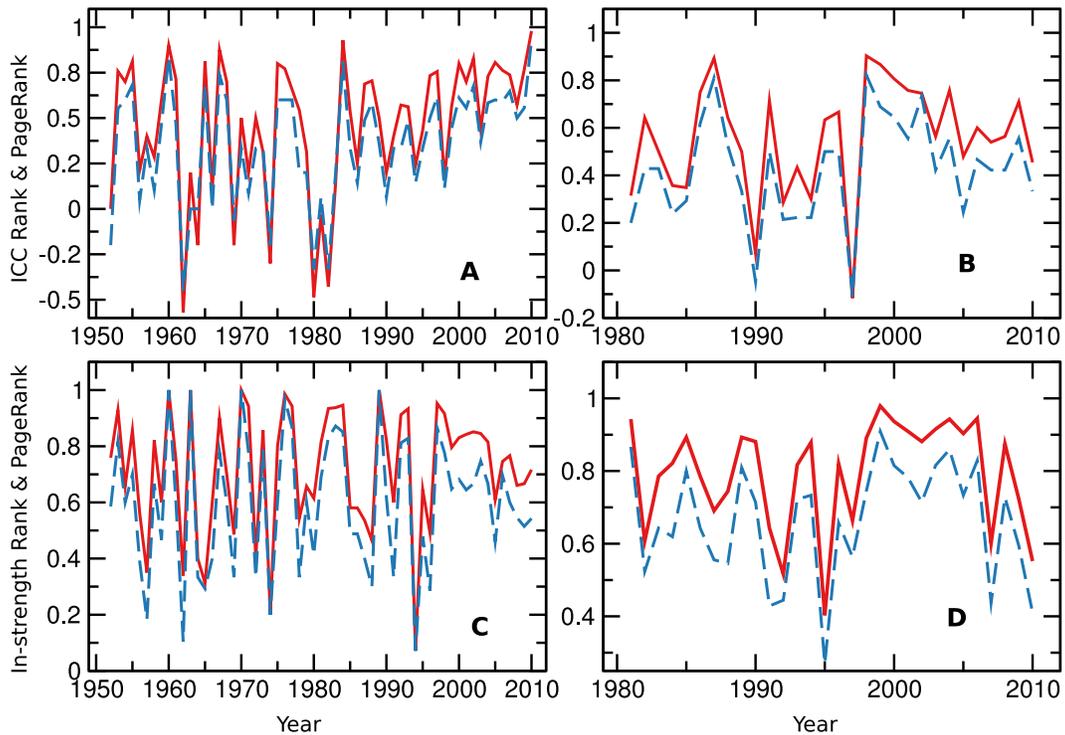}
\end{center}
\caption{{\bf Correlation among different ranking schemes.} {\bf (A)} Spearman correlation coefficient (red) and Kendall $\tau$ (blue), between the ranking based on PageRank and the one based on the Reliance ICC Ranking, as function of time, for Test matches ($1952-2010$).  {\bf (B)} The correlation coefficients are calculated between the ranking based on PageRank and the one Reliance ICC Ranking for ODI matches ($1981-2010$). {\bf (C)} The correlation coefficients are calculated between the ranking based on PageRank and In-strength for Test matches ($1952-2010$).  {\bf (D)} The correlation coefficients are calculated between the ranking based on PageRank and In-strength for ODI matches ($1981-2010$).}
\label{fig5}
\end{figure}

\begin{table}
\centering
\caption{{\bf Ranking of captains in different era in Test history.} We have shown the ranking of top five captains between $1877 - 2010$ as well as their nationality. A network of competing captains are generated for each era. We run the ranking procedure and rank the captains according to their PageRank score.}
\begin{tabular}{ccc}

\toprule
\multirow{2}{*}{}
 \textbf{Era} & \textbf{Top five captains}  & \textbf{Country}\\ 
\midrule 
\multirow{4}{*}{\textbf{1877-1950}}	
	& \multirow{4}{*}{}
	
	Bill Woodfull & Australia      \\ 
	& Sir Donald Bradman & Australia      \\
	& John Goddard & West Indies   \\ 
	& Sir Gubby Allen & England     \\
        & Normal Yardley & England   \\ 
\midrule 

\multirow{4}{*}{\textbf{1951-1960}}	
	& \multirow{4}{*}{}
	
	Richie Benaud & Australia      \\ 
	& Gulabrai Ramchand & India        \\
	& Peter May & England  \\ 
	& Abdul Kardar & Pakistan     \\
        & Lindsay Hassett & Australia   \\ 

\midrule 

\multirow{4}{*}{\textbf{1961-1970}}	
	& \multirow{4}{*}{}
	
	Richie Benaud & Australia     \\ 
	& Sir Frank Worrell & West Indies       \\
	& Bob Simpson & Australia      \\
	& Ted Dexter & England   \\ 
	& Sir Garry Sobers & West Indies     \\
 
\midrule 

\multirow{4}{*}{\textbf{1971-1980}}	
	& \multirow{4}{*}{}
	
	Ian Chappel & Australia      \\ 
	& Clive Lloyd & West Indies        \\
	& Greg Chappell & Australia   \\ 
	& Ray Illingworth & England     \\
        & Mike Denness & England    \\ 

\midrule 

\multirow{4}{*}{\textbf{1981-1990}}	
	& \multirow{4}{*}{}
	
	Sir Vivian Richards & West Indies      \\ 
	& Allan Border & Australia        \\
	& Greg Chappell & Australia   \\ 
	& Clive Lloyd & West Indies     \\
        & Geoff Howarth & New Zealand    \\ 

\midrule 

\multirow{4}{*}{\textbf{1991-2000}}	
	& \multirow{4}{*}{}
	
	Mark Taylor & Australia      \\

	& Hansie Cronje & South Africa        \\
	& Allan Border & Australia   \\ 
	& Mike Atherton & England     \\
        & Steve Waugh & Australia    \\ 

\midrule 

\multirow{4}{*}{\textbf{2001-2010}}	
	& \multirow{4}{*}{}
	
	Graeme Smith & South Africa      \\ 
	& Ricky Ponting & Australia        \\
	& Steve Waugh & Australia   \\ 
	& M. S. Dhoni & India     \\
        & Sourav Ganguly & India    \\ 

\bottomrule
\end{tabular}
\label{table5c}
\end{table}

\begin{table}
\centering
\caption{{\bf Ranking of captains in different era in ODI history.} A network of teams is generated for each era. We then run the PageRank algorithm on each network which gives a PageRank score. The teams are then ranked according to their PageRank score. We have shown the ranking of top five captains between $1971 - 2010$ as well as their nationality.}
\begin{tabular}{ccc}

\toprule
\multirow{2}{*}{}
 \textbf{Era} & \textbf{Top five captains}  & \textbf{Country}\\ 
\midrule 
\multirow{4}{*}{\textbf{1971-1980}}	
	& \multirow{4}{*}{}
	
	Greg Chappell & Australia      \\ 
	& Clive Lloyd & West Indies        \\
	& Geoff Howarth & New Zealand   \\ 
	& Mike Brearley & England     \\
        & Sunil Gavaskar & India    \\ 
\midrule 

\multirow{4}{*}{\textbf{1981-1990}}	
	& \multirow{4}{*}{}
	
	Imran Khan & Pakistan      \\ 
	& Sir Vivian Richards & West Indies        \\
	& Kapil Dev & India  \\ 
	& Allan Border & Australia     \\
        & Javded Miandad & Pakistan   \\ 

\midrule 

\multirow{4}{*}{\textbf{1991-2000}}	
	& \multirow{4}{*}{}
	
	Hansie Cronje & South Africa     \\ 
	& Arjuna Ranatunga & Sri Lanka       \\
	& Mohammad Azharuddin & India       \\
	& Wasim Akram & Pakistan   \\ 
	& Richie Richardson & West Indies     \\
 
\midrule 

\multirow{4}{*}{\textbf{2001-2010}}	
	& \multirow{4}{*}{}
	
	Ricky Ponting & Australia      \\ 
	& Graeme Smith & South Africa        \\
	& M. S. Dhoni & India   \\ 
	& Stephen Fleming & New Zealand     \\
        & Mahela Jayawardene & Sri Lanka    \\ 

\bottomrule
\end{tabular}
\label{table6b}
\end{table}

\begin{figure}[!ht]
\begin{center}
\includegraphics[width=5.5in]{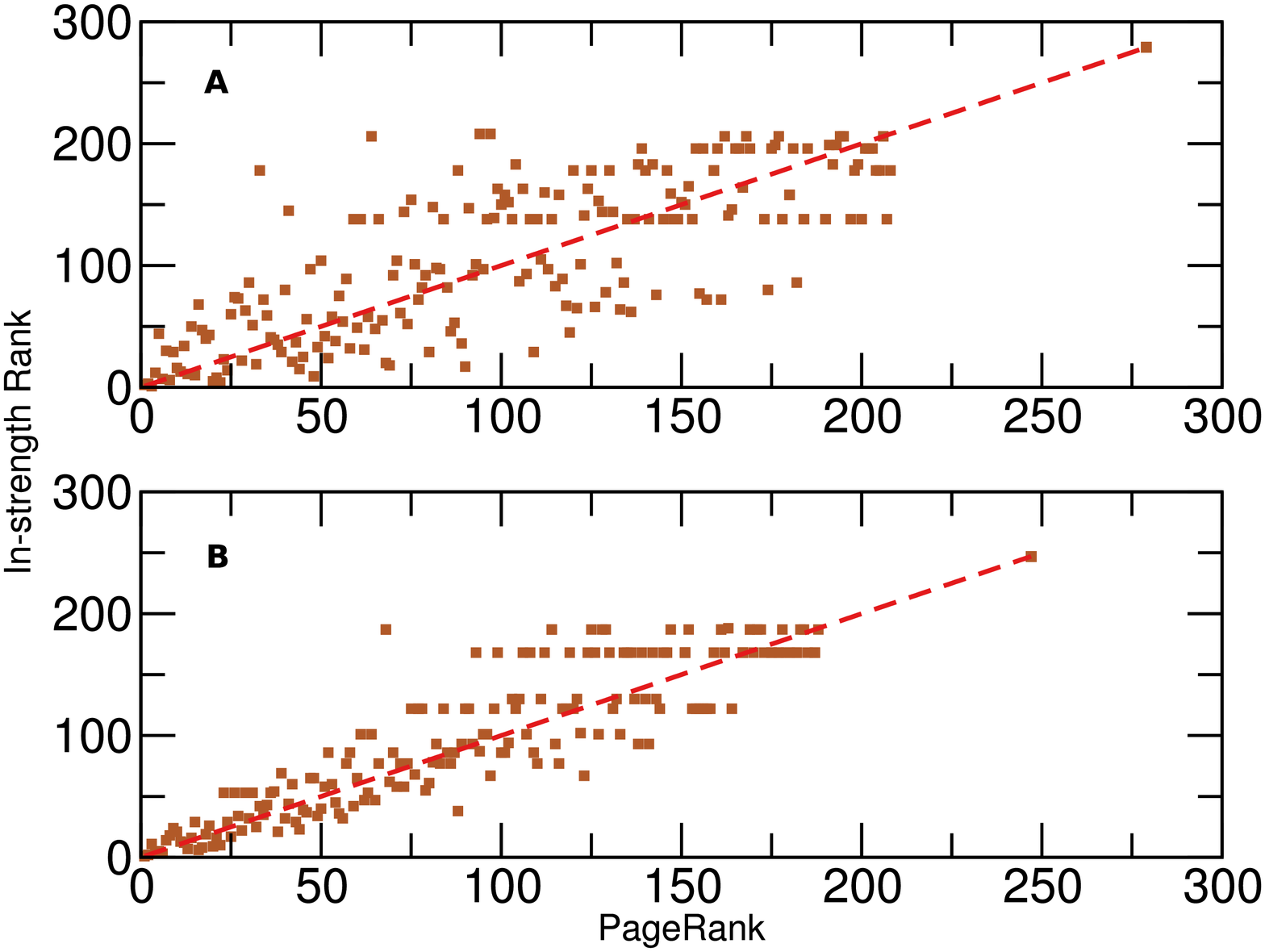}
\end{center}
\caption{{\bf Relation between PageRank and in-strength Rank for captains. } {\bf (A)} Scatter plot between the rank positions obtained according to in-strength and those obtained with PageRank for Test cricket ($1952-2010$); (Kendall ${\tau}=0.734$, Spearman correlation ${\rho}=0.892$). {\bf (B)} Scatter plot between the rank positions obtained according to in-strength and those obtained with PageRank for ODI cricket ($1981-2010$); (${\tau}=0.836$, ${\rho}=0.948$). }
\label{fig6}
\end{figure}

\section{Conclusion}

Our work demonstrates the strength of social network analysis methods in quantifying the success of cricket teams and their captains. Here we have created a directed and weighted network of contacts (i.e, teams and captains). The correct assessment of a team's success (or captain's success) needs the consideration of the entire network of interaction. The PageRank algorithm takes into account the quality of matches won. For example, a win against a strong team is more important than a win against a weak team. Also a captain is as good as the team. In this sense, a win against {\it Clive Lloyd}, {\it Steve Waugh} or {\it Graeme Smith} is more relevant than a win against a lesser captain. Our analysis shows that PageRank algorithm is effective in finding the most successful team and captain in the history of cricket.
\newline
It should be noted that success of a team or a captain depends on various factors like home advantage, success of batsmen and bowlers. For example, Australia's dominance in both forms of the game is a manifestation of the fact that they are able to adjust in all kinds of pitches around the world, whereas subcontinent teams always played well under subcontinent conditions but were not able to repeat their performance abroad on a consistent basis. Our analysis does not require these `external' factors which are usually taken into account in ICC rankings. However, we would like to mention that our method does not aim to replace the ICC ranking. It suggests a novel approach to refine the existing ranking scheme.
\newline
We would like to state that cricket is a team game. Success or failure of a team depends on the collective performance of all team members. Simple statistics like runs scored by batsmen, wickets taken by bowlers or exceptional fielding does not provide a reliable measure of a player's contribution to the team's cause. Quantifying the impact of player's individual performance in sports has been a topic of interest in soccer (\cite{duch10}) and baseball (\cite{saavedra09}). However, in cricket the rules of the game are different and therefore it would be interesting to apply tools of network analysis on interaction between players. For example, a contact network of batsman $vs.$ bowler could give an estimate of the greatest batsman (bowler) ever. Potentially, a quantitative approach to a player's performance could be used to estimate the Man of the Match (Series) award after a tournament.

\section*{Acknowledgement}
We thank the cricinfo website for the public availability of information on cricket matches. We also gratefully acknowledge helpful discussions with Rufaro Mukogo, David Mertens and Xiaohan Zeng. 

\end{document}